\documentclass[10pt,letterpaper]{article}
\usepackage{opex3}
\usepackage{color}
\usepackage{placeins}
\usepackage[colorlinks=true,urlcolor=blue,linkcolor=black,citecolor=black,breaklinks=true,bookmarks=false]{hyperref}
\usepackage{epstopdf}
\usepackage{verbatim}

\begin{document}

\title{Anomalous dispersion in Lithium Niobate one-dimensional waveguide array in the near-infrared wavelength range}

\author{Alin Marian Apetrei$^1$ Alicia Petronela Rambu$^1$ Christophe Minot$^{2,3}$ Jean-Marie Moison$^2$ Nadia Belabas$^2$ and Sorin Tascu$^{1,*}$}

\address{$^1$Research Center on Advanced Materials and Technologies, Sciences Department, "Alexandru Ioan Cuza" University of Iasi, 11, Boulevard Carol I, 700506 Iasi, Romania\\
$^2$Laboratoire de Photonique et de Nanostructures, Centre National de la Recherche Scientifique - UPR20, Route de Nozay, 91400 Marcoussis, France\\
$^3$Institut Mines-Telecom, Telecom ParisTech, 46 rue Barrault, 75634 Paris Cedex 13, France}

\email{$^*$sorin.tascu@uaic.ro} %% email address is required

\begin{abstract}
Knowing the dispersion regime (normal vs anomalous) is important for both an isolated waveguide and a waveguide array. We investigate by the Finite Element Method the dispersion properties of a $LiNbO_3$ waveguides array using two techniques. The first one assumes the Coupled Mode Theory in a 2-waveguide system. The other one uses the actual diffraction curve determined in a 7-waveguide system. In both approaches we find that by decreasing the array period, one passes from normal dispersion by achromatic point to anomalous array dispersion. We then illustrate the wavelength separation by doing Runge-Kutta light propagation simulations in waveguide array. As all the parameters' values are technologically feasible, this opens new possibilities for optical data processing.
\end{abstract}


\section{Introduction}
This work deals with the dispersive properties of lithium niobate $LiNbO_3$ (LN) coupled waveguides array (WA).

\par{On one hand, coupled waveguide arrays are a promising tool in photonics in general and in optical data processing in particular. This can be implemented either by making heterogenous structures (i.e. having several zones with different coupling characteristics between the waveguides within each zone [1,2]), or by use of nonlinear optical effects, especially the Kerr effect that makes possible self focusing [3] and beam routing [4,5].}

\par{On the other hand, LN is already one of the most used material for many integrated optical applications and devices. The maturity of the technological process, the quality of fabricated structures (e.g. waveguides [6,7]) and the excellent electro-optical and nonlinear optical properties of the material are exploited in various applications such as laser frequency doublers, wideband tunable light sources, light amplification, quasi-phase-matched frequency convertors, information and image storage, surface acoustic wave, optical switches, optical modulators, multiplexors, and for all-optical processing of signals devices. The extensive literature available on LN single crystal applications and devices shows the wide interest in the areas of both numerical investigation and applied research.}

\par{That is why combining LN and WA seems a natural choice. The literature on LN coupled waveguides include several research directions, such as discrete spatial solitons \cite{8,9}, discrete diffraction by using electro-optic effect in periodically poled LN (PPLN) structures \cite{10}, optical switching and power control \cite{11} and high dimensional quantum states \cite{12}.}

\par{Dispersion is one of the most important optical properties of any material or device with potential application in photonics (e.g. wavelength separation, soliton formation, etc... ). In analogy with bulk material where the propagation governing parameter is the refractive index, the propagation in coupled waveguides arrays is mainly governed by the coupling constant $C$. The analogy may continue: both the refractive index and coupling constant depend on wavelength.}

\par{ Our bibliographic study reveals there is a lack of investigation of dispersion properties in coupled LN waveguides. In this context, the study of dispersion properties in this system could open the route for a breakthrough in photonics. Therefore, in this study we numerically investigate the dispersive properties of homogeneous array of LN waveguides.}

\section{Theoretical aspects}
In an isolated waveguide, electromagnetic waves have a constant amplitude and spatial periodicity defined by the propagation constant $\beta_0$. In the presence of a neighbor waveguide, the light is evanescently coupled to this one with a certain "strength", known as the coupling constant $C$. $C$ depends on the waveguide separation $s$ and wavelength $\lambda$ and is related to the spatial overlap of the electric field distributions of the two waveguides. The coupling results in periodically varying amplitudes. This corresponds to an equivalent propagation constant $K_z>>\beta_0$.

\par{This formalism can be extended to the propagation through an array of $N$ waveguides. It is the Floquet-Bloch (FB) formalism. It states that light propagation can be described by large collective waves, with modal amplitudes $A_i$ across the array $i=1,N$ and a wavevector whose components are $K_z$ and $K_x$ ($x$ along the direction perpendicular to waveguides). Between them, there is a very important relation, known as the diffraction relation. For an 1-D array of $N$ waveguides, there are $N$ collective FB waves. The coupling can be modified via the distance between the waveguides}.

\par{In the \textbf{"weak"} coupling regime, the coupling between waveguides is described by the Coupled Mode Theory (CMT), where only the coupling with the first neighbor is considered and there is no deviation from the orthogonality of the modes. The only parameter one needs to characterize the coupling is the coupling constant $C$ between two waveguides only. This can be analytically or numerically calculated (depending on the complexity of the system) either :}

 \begin{itemize}
   \item \textit{(i)}   as the overlap integral of the field distributions $C\propto\int\overrightarrow{E}_{0i}(x,y)\overrightarrow{E}_{0j}(x,y)dxdy$ of two isolated waveguides ;
   \item \textit{(ii)}  or by taking a two-waveguide system and obtaining the two collective modes, having the propagation constants $\beta_S$ and $\beta_A$, where the indices A and S stand for anti-symmetric and symmetric modes respectively. Then, one computes $C=(\beta_S-\beta_A)/2$.
 \end{itemize}

The FB waves $exp(iK_zz+iK_xx)$ obey the diffraction relation :

\begin{equation}
K_z(K_x)-\beta_0=2Ccos(K_xs)
\label{eq1}
\end{equation}
where $s$ is the array period and $\beta_0$ is the propagation constant of the isolated waveguide. The diffraction relation controls the behavior of limited-size beams. The slope $P=X/Z$ of the maximum intensity trajectory (beam direction) is given by the first derivative of Eq.(1), as :

\begin{equation}
P=-dK_z/dK_x
\label{eq2}
\end{equation}

and the second derivative gives the divergence of the beam.

\par{There is an important theoretical and practical case at $K_x=\pi/2s$. The slope of this beam becomes :}

\begin{equation}
P^{CMT}=2Cs
\label{eq3}
\end{equation}

and the second derivative is zero. This means that for this particular beam the divergence is null. As one can see, $P$ depends on $C$ so it depends on wavelength. By deriving $dC/d\lambda$ one obtains the slope's derivative :

\begin{equation}
 dP^{CMT}/d\lambda=2*s*dC/d\lambda.
 \label{eq4}
\end{equation}

In analogy with a wavelength separation device (e.g. a prism), the slope's derivative gives the \textit{angular dispersion of the array}.

\par{In the case of a \textbf{"stronger"} coupling, the second order contribution must be taken into account and the extended CMT theory [13, 14] is required. The diffraction relation becomes :}

\begin{equation}
K_z(K_x)-\beta_0=2C\frac{cos(K_xs)+\xi+2\zeta cos(2K_xs)}{1+2\eta cos(K_xs)}
\label{eq5}
\end{equation}

where $\eta$, $\zeta$ and $\xi$ are correction factors. $\eta$ takes into account the non-orthogonality of the modes and $\zeta$ and $\xi$ the second-order neighbor coupling. If all three factors vanish, the dispersion relation simplifies to the weak coupled case. So, all of them depend on the period. For $K_x=\pi/2s$, the slope of the beam is :

\begin{equation}
P^{ eCMT}=2Cs[1+\eta(2\zeta-\xi)]
\label{eq6}
\end{equation}

\par{For shallow-ridge GaAs/GaAlAs and InP/InGaAsP waveguides arrays, the literature reports that $\eta>\zeta>\xi$ [14]. The adjacent question this study tries to respond also, is wether the same qualitative relation stands for LN waveguide arrays as well or not. We will investigate this aspect in the next section. Would this be the case, this means the slope in eCMT is higher than in classical CMT, $P^{eCMT}>P^{CMT}$, as if the coupling constant $C$ was higher. It might be seen as an apparent coupling constant : }

\begin{equation}
C^{app}=C(1+\eta(2\zeta-\xi))
\label{eq7}
\end{equation}

\par{The main goal of this work is to evaluate the angular dispersion $dP/d\lambda$ of the WA. In order to illustrate how wavelength separation works in this system based on the dependence of the beam's slope on the wavelength, we then do Runge-Kutta numerical simulations of light propagation in the waveguide array.}

\section{Study of dispersion in two-waveguide systems and multi-waveguide systems}

Our approach is two-fold, as follows :

\begin{itemize}
  \item An approximative one, assuming a CMT cosine diffraction relation. We first numerically calculate the propagation constants of a two-waveguide system. Then we determine the coupling constant $C$  using method \emph{(ii)} for the reason it takes into account the non-orthogonality of the modes as well. We vary the array period $s$ and wavelength $\lambda$ and thus obtain $C(s,\lambda)$. Then we obtain $dC/d\lambda$ and thus $dP/d\lambda$ function of $s$.
  \item The accurate approach, using the actual diffraction relation. We compute the collective modes of a 7-waveguide array. So, we have the (sampled) band structure $K_z(K_x)$. Then we compute the derivative $-dK_z/dK_x$ at $K_x=\pi/2s$ and thus obtain the slope $P$ of the (quasi)non divergent beam. Repeating the calculus for different sets of $\lambda$ and $s$, one obtains $dP/d\lambda$ as a function of $s$.
\end{itemize}

\subsection{Technical details : index profile and Finite Element Method (FEM) calculations}

We use Finite the Element Method (FEM) based Comsol software to retrieve the eigen modes of both LN two-waveguide and multi-waveguide systems. The structure we simulate is a LN isotropic rectangle with an air slab on top of it. The refractive index spatial profile of the isolated waveguide is :

\begin{equation}
n(x,y)=n_{LN}+\Delta n*e^{\frac{y}{w_y}}*[\prod(-\infty, \frac{-d}{2})e^{(\frac{x+d/2}{w_x})^2}+\prod(\frac{-d}{2},\frac{d}{2})+\prod(\frac{d}{2},\infty)e^{(\frac{x-d/2}{w_x})^2}]
\label{eq8}
\end{equation}

$\prod(x_1,x_2)$ is the rectangular function. The waveguides have typical characteristics that can be fabricated by Soft Proton Exchange (SPE) technique. If different Proton Exchange techniques used to create waveguides may reduce or destroy the nonlinear coefficient in LN and/or periodic domain orientation in PPLN substrate, the SPE has the advantage of preserving both the nonlinear coefficient and the domain orientations. With this technique, only TM modes are guided modes. We have to consider the air slab into account because, by this technique, the core of the waveguides is a few microns only below the surface [15]. Thus, for $y>0$ the air slab ensures the vertical confinement, whereas for $y<0$ the local increase of the refractive index that plays the role.

\par{We describe the values chosen for each parameter :}

 \begin{itemize}
 \item $n_{LN}$ is the refractive index of bulk LN that takes into account the material dispersion through the Selmmeier relation;
 \item $w_y=2.1 \mu m $ is the depth at 1/e of the exponential profile in the $y$ direction. It has to be sufficiently high for good mode confinement, but not too high, in order to avoid the appearance of higher order modes. Anyway, in practice, $w_y$ can hardly be increased significantly for reasonably processing times (several days);
 \item $\Delta n=0.022 $ is the index contrast. We took this value because it is in the upper limit range of SPE. In practice, $\Delta n$ too, can hardly be increased significantly by SPE;
 \item $d=2 \mu m $ is the waveguide width. We took into account several values for $d$ and finally chose this particular value because it is a good compromise between several technological and theoretical requirements. A narrow waveguide is good for having a quasi-monomode behavior in a large wavelength range (0.98 - 1.55 $\mu m$). It also ensures a high power density, if ever needed, for an increased nonlinear optical efficiency. A wide waveguide on the other hand ensures low insertion loss and lower propagation loss through the scattering at the interfaces;
 \item $w_x=0.4 \mu m$ is a small broadening of the index profile in the direction of the coupling between adjacent waveguides, due to diffusion.
 \end{itemize}

The index profile is then two-fold or seven-fold replicated with the period $s$, where $s$ is center-to-center distance between adjacent waveguides.

\subsection{The two-waveguide system : coarse evaluation}

\begin{figure}[h]
\centering\includegraphics[width=13cm]{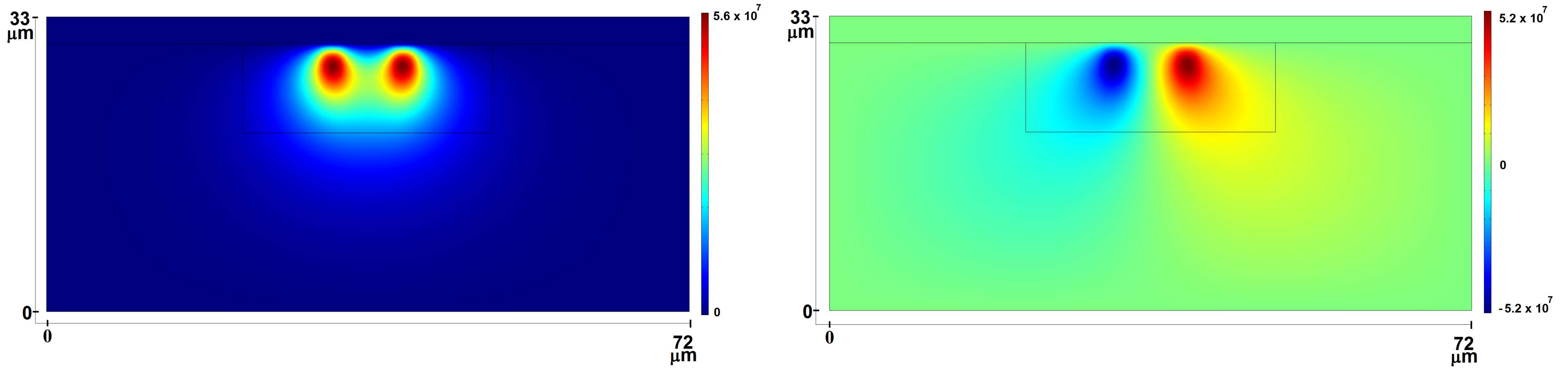}
\caption{(Color online) Electric field spatial distributions (linear scale) of the symmetric (left) and antisymmetric (right) modes of a two identical waveguides with $s=8 \mu m$ at $\lambda=1.55 \mu m$. The small rectangle around the intensity maxima has no physical significance. It is only used for increasing mesh density (space discretization).}
\label{fig1}
\end{figure}

In Figure \ref{fig1} we present, as an example, the electric field profiles of the two modes at $\lambda=1.55 \mu m$ for $s=8 \mu m$ (in linear scale).

\FloatBarrier

\begin{figure}[h]
\centering\includegraphics[width=12cm]{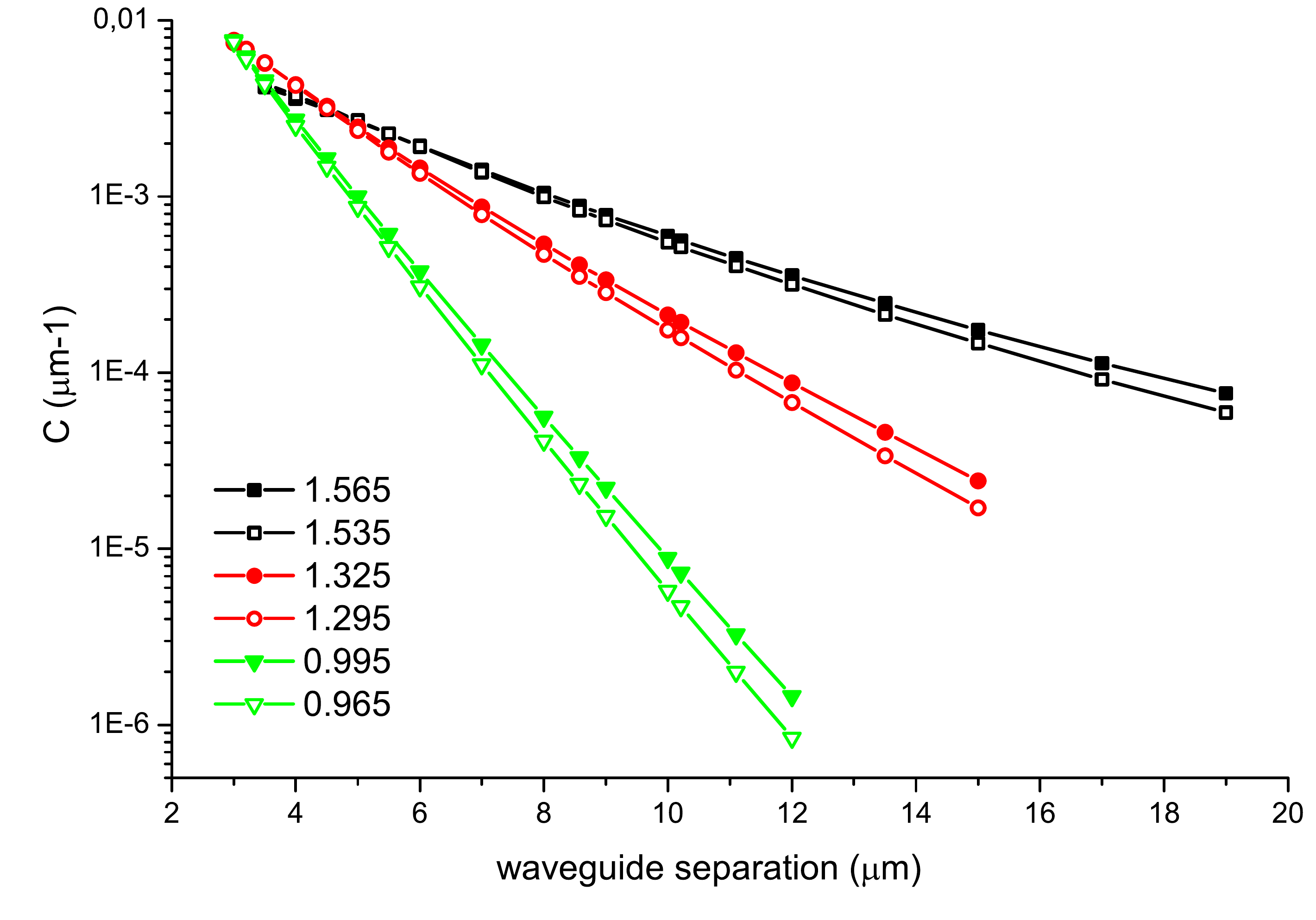}
\caption{(Color online) Coupling constant $C=\pi(n_a-n_s)/\lambda$ as function of the array period for wavelengths 1535 nm (filled squares), 1565 nm (open squares), 1325 nm (filled circles), 1295 nm (open circles), 995 nm (filled triangles), 965 nm (open triangles).}
\label{fig2}
\end{figure}

 For the 2-waveguide case, the Comsol simulations are very robust with respect to boundary conditions, because the modes are well confined. We can simulate a large structure, the border being far of the mode center. The structures with and without Perfectly Matched Layers (PML) give effective indices that differ insignificantly for this study ($\sim10^{-7}$). Then, we use the propagation constants to obtain $C$ as $(\beta_S-\beta_A)/2$. Figure \ref{fig2} presents $C$ as function of period for 3 sets of wavelengths, i.e. 965 and 995 nm (around the central wavelength $\lambda_c=980 nm$, not presented), 1295 and 1325 nm ($\lambda_c=1310 nm$), 1535 and 1565 nm ($\lambda_c=1550 nm$). The curves for wavelengths +- 15 nm apart from the central wavelengths allow us to find the derivative of the slope, $dP/d\lambda$.

\FloatBarrier

\par{The coupling constant depends as expected nearly exponentially on the period. The resulting slope $P$ and its derivative $dP/d\lambda$ figures for the 2-waveguide system are presented later on.}

\subsection{The seven-waveguide system : accurate evaluation}

For the 7-waveguide system, the FEM investigation is less straightforward and more difficult. The high order modes 6 and 7 are more sensitive to boundary conditions. It should not be the case for an infinitely large structure. In order to minimize the border influence, we tried different boundary conditions, such as with and without PML. Due to asymmetry in the structure (the air slab is present on the top side of the LN only), the PML boundary conditions, that must be symmetrical, are not the best option here. That is why we use Scattering Boundary Condition (SBC) that is adequate for our asymmetric structure and, in the same time, presents the advantages of PML (eliminate the parasite border reflections and remove the leaking/evanscent field touching it) and it is acceptable in terms of hardware resources needed. More than this, we used fillet edge corners in order for the wavefront to touch as much as (qualitatively) possible at normal incidence the boundary for all 7 modes. This demand is imposed by the SBC. The 7 eigen-modes give the diffraction curve.

\par
\begin{figure}[h]
\includegraphics[width=12cm]{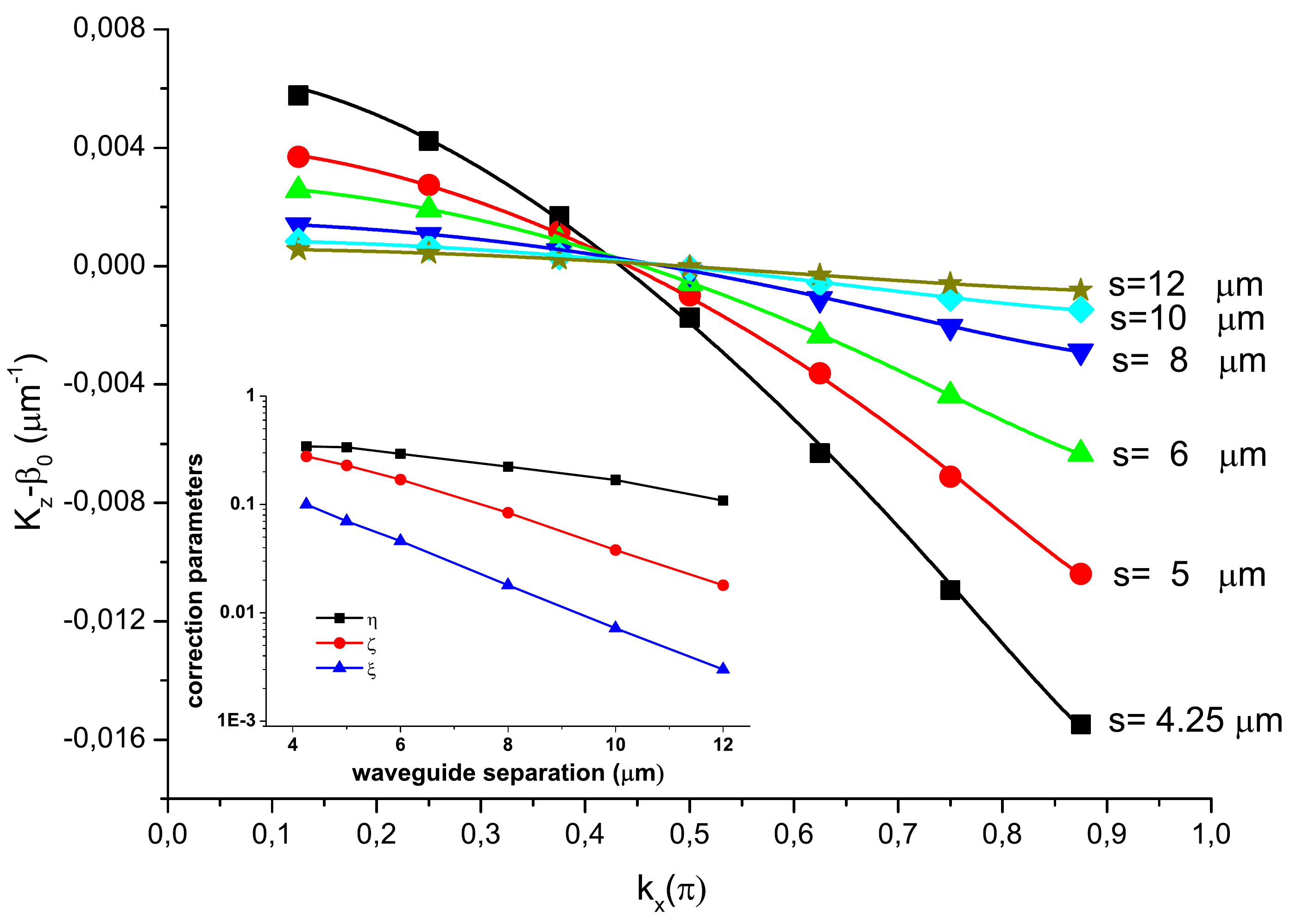}
\caption{(Color online) Points: diffraction bands of a 7 waveguide system at $ \lambda=1.55 \mu m$ for different periods $s$ (given in the legend, in $\mu m$). The error bars are smaller than the symbols. Lines: fit with the equation \ref{eq5}. Inset: Correction parameters $\eta, \zeta, \xi $ function of the period at $\lambda=1.55 \mu m$}
\label{fig3}
\end{figure}

{Figure \ref{fig3} presents a set of discrete diffraction curves $K_z(K_x)$ at $\lambda = 1.55 \mu m$, with the period $s$ as parameter. The abscissa is in reduced units, i.e. $k_x=K_xs$ in order to represent all the curves on the same graph. Line curves are fitted with equation \ref{eq5}.}

\FloatBarrier

As one can see, the curves drift more and more from the cosine curves as the period decreases. For the readability of the figure, we did not put the error bar because these are smaller than the symbols (relative error is less than $1.5\% $), even for the worst case, i.e. the 7-th mode for the closest waveguides ($s=4.25 \mu m$). Nevertheless, we make the remark that for the slope and, especially for the slope's derivative, the relative errors are bigger, as the slope $P$ of the beam involves the first derivative of the diffraction curve and the derivative of the slope $dP/d\lambda$ involves second order derivative.

\par{From the fit of these curves with the equation \ref{eq5}, we obtained the correction parameters $\eta,  \zeta,  \xi $. We check the quality of the fit by the statistical parameter adjusted R-square ($\overline{R}^2$). This is an improved version of the R-square ($R^2$) for the number of variables in a model. In contrast with ($R^2$) that only increases with every new variable added to the model, the adjusted R-squared compensates for the addition of variables and only increases if the new term enhances the model above what would be obtained by chance. For all the fitted curves, $\overline{R}^2>0.9988$. }

\par{The dependence of the correction parameters with the period is presented in the inset of the Figure \ref{fig3}. As one can see, these factors obey the same relation $\eta>\zeta>\xi$ as for shallow-ridge GaAs/GaAlAs and InP/InGaAsP waveguides arrays [14]. As a consequence of this result, we expect for the slope within eCMT to be higher that in CMT. This aspect is investigated in the next subsection.}

\subsection{Beam slope evaluation}

For the two-waveguide system the slope calculation is straightforward, as in eq. \ref{eq3}. For the seven-waveguide system the slope calculation is as in eq. \ref{eq2}, but it is less straightforward. We tried several methods. They all qualitatively agree. We finally chose the most robust one. This is the direct calculation, i.e. starting from the 7 points discrete dispersion curve. We calculate the derivation in the $k_x=\pi/2$ point of the dispersion band, i.e. around the $4^{th}$ mode. The derivative is the average of the left and the right derivatives. This implies the $K_z$ of the $3^{rd}$, $4^{th}$ and the $5^{th}$ modes only. As the absolute error of the $5^{th}$ mode effective refractive index is relatively small $(\sim10^{-5})$ and those of the $3^{rd}$ and $4^{th}$ modes are negligible $(<10^{-6})$, this method proved to be the most robust of all.

\par
\begin{figure}[h]
\includegraphics[width=12cm]{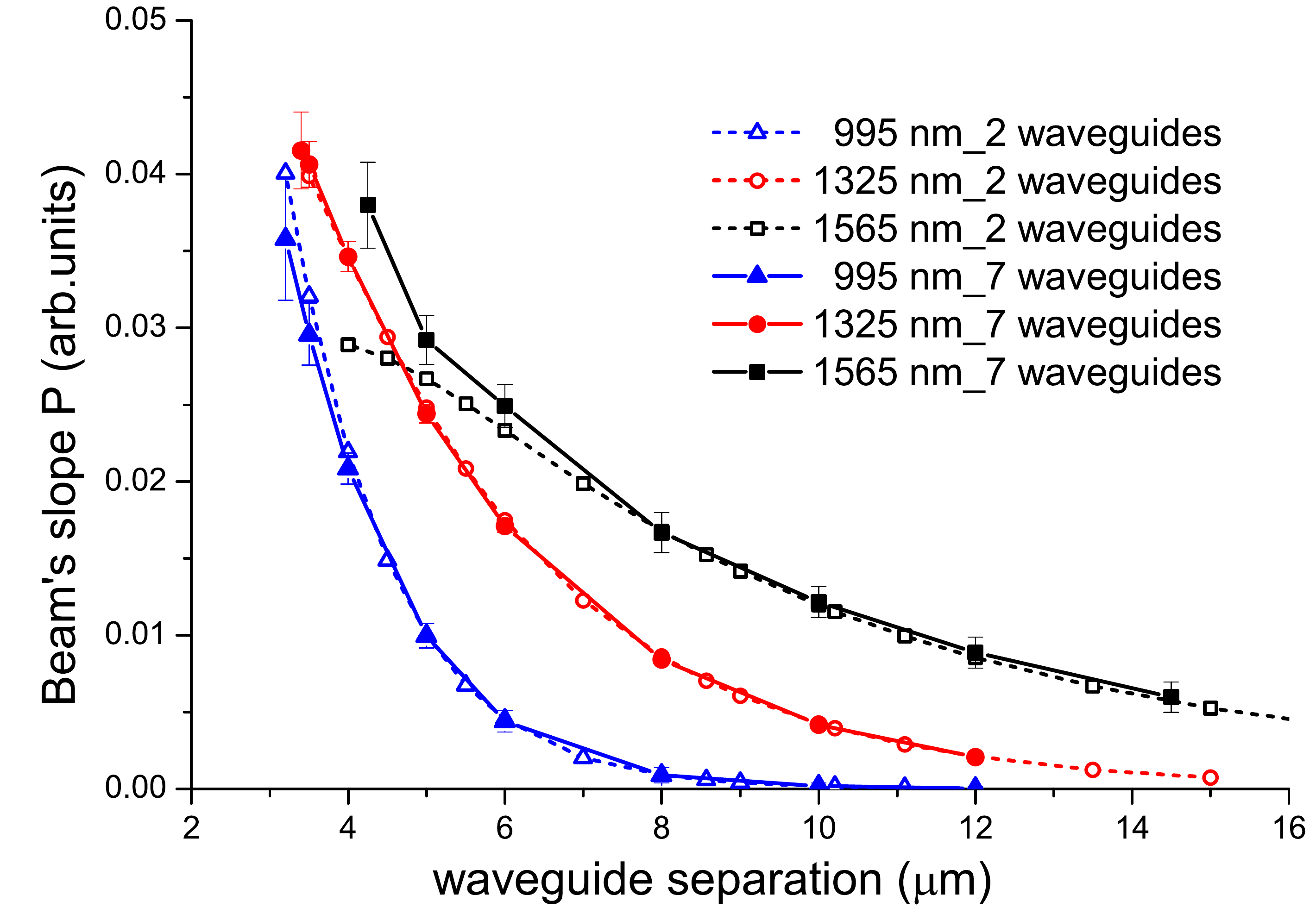}
\caption{(Color online) Slope of the beam obtained in the two-waveguide system CMT approximation (open symbols dotted line) and in the 7-waveguide system eCMT approximation (filled symbols plain lines) at $K_x=\pi/2s$ (non-divergent beam) as a function of the period $s$. The parameter is the wavelength ($\lambda = 1.565 nm$ squares, $\lambda = 1.325 nm$ circles, $\lambda = 995 nm$ triangles).}
\label{fig4}
\end{figure}

{In fig. \ref{fig4} we present the beams' slopes $P$ at $980nm, 1310 nm $ and $1550 nm$ obtained in these two evaluations.}

\FloatBarrier

For wavelengths where the coupling is strong, i.e. at $\lambda=1.55 \mu m$, the error bar is small enough for the results to be eloquent, namely, in most cases, the two evaluations are in good agreement. A significant difference occurs only at high wavelengths and very strong couplings, where the slope within accurate evaluation is higher than in the coarse one (as expected from the theory presented above and anticipated in the end of previous subsection). For shorter wavelengths, the error bar becomes higher and the results are blurred.

\subsection{Angular dispersion : slope's dispersion }

\begin{figure}[h]
\includegraphics[width=12cm]{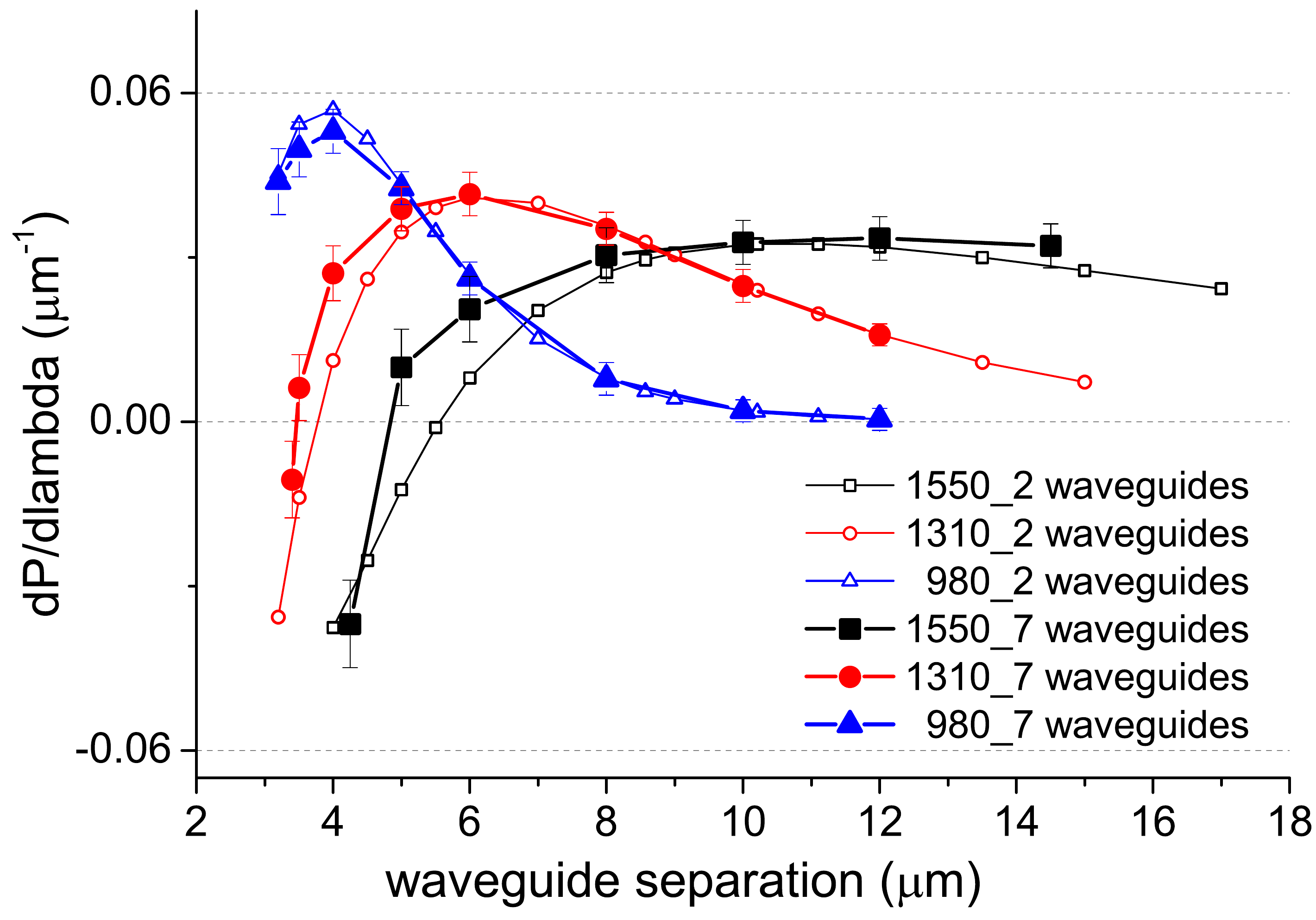}
\caption{(Color online) Derivative of slope of the FB beam obtained in the two-waveguide system (open symbols) and in the 7 waveguide system (filled symbols) at $k_x=\pi/2$ (non-divergent beam) function of the period $s$. The parameter is the wavelength ($\lambda = 1550 nm, 1310 nm, 980 nm$). }
\label{fig5}
\end{figure}

 For the two-waveguide system, as presented in the previous subsections, for each central wavelength $\lambda_c$ is 980 nm, 1310 nm and 1550 nm we obtained two additional curves $C$ for wavelength separated by 15 nm and retrieved $dC/d\lambda$. As for the seven-waveguide system, we obtained the slopes $P$ (as in the previous subsection) for the same wavelength $\pm15 nm$ around the central wavelength. Figure \ref{fig5} presents the derivative of the slope of the beam (the angular dispersion of the array) obtained within the two evaluations.

\FloatBarrier

The most important conclusion is that decreasing the period, one passes from a normal dispersion regime where beam's slope increases with wavelength, through achromatic point, to an anomalous array dispersion, where the beam slope decreases with wavelength.

\par{We emphasize that it is not related to the bulk (material) dispersion, because, in this wavelength range, the LN is in normal dispersion regime, up to $\lambda=1.9 \mu m$, as results from ref. [16]. It is the array (collective) dispersion that is anomalous. More important, it can be tuned by modifying the period. This behavior is qualitatively similar to the waveguide dispersion encountered in some  microstructured optical fibers [17,18]. For these fibers with very small effective mode area [19,20] the zero-dispersion-wavelength can be tuned in a wide range by changing the geometrical parameters, such as the pitch $\Lambda$, hole-diameter over pith ratio ($d/\Lambda$) and the numbers of hole rings .}

\par{We note the qualitative agrement between the two different evaluations. Both of them predict the existence of the achromatic point and of the anomalous dispersion regime.}

\section{Wavelength separation demonstrator}

As an illustration of the dispersive properties of such a system, we solve the N-coupled equations governing the propagation through the array. The propagation is governed by the equation:

\begin{equation}
    \frac{\partial a_m}{\partial z} = i(C^{app}_{m+1,m}a_{m+1}+C^{app}_{m,m-1}a_{m-1})
    \label{eg9}
\end{equation}
where $a_m$ is the complex amplitude in the $m$-waveguide.

\par{We used a hybrid approach. It consists of formally taking into account the coupling with first neighbor only (for the simplicity of the calculus sake), but with the apparent coupling constant $C^{app}$ that takes into account the increasing of the coupling constant because of the second-order neighbor influence and the non-orthogonality of the modes (see equation \ref{eq7}).}

\par{We use Runge-Kutta method (of order five minimum) from the IMSL commercially available numerical library. The code was verified by testing it on simple cases that have analytical solutions, such as the input in one waveguide only. The solutions are in this case the Bessel J functions. This simple numerical case is well treated in the literature. }

\par
\begin{figure}[h]
\includegraphics[width=12cm]{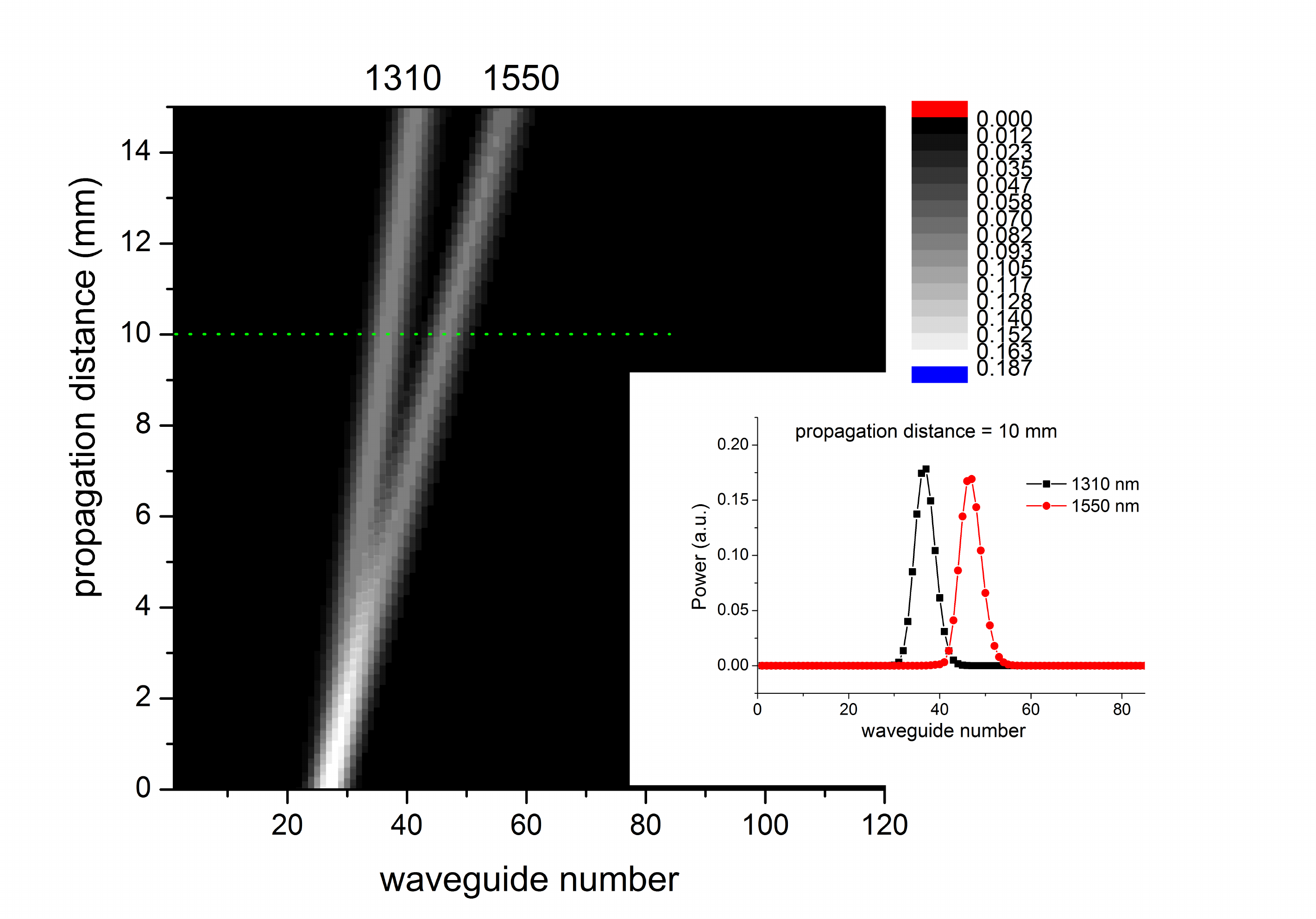}
\caption{(Color online) Top view of the spatial evolution of the power (square of the amplitude) through the array of 120 waveguides for two beams (1310 nm and 1550 nm). The input beams have a gaussian spatial distribution with waist = 5 array period. Abscise is the waveguide number. The inset is the profile of the two beams after 10 mm propagation.}
\label{fig6}
\end{figure}

{Figure \ref{fig6} presents the spatial evolution of two gaussian beams with input waist = 5$s$, where $s=8 \mu m$, hence in normal dispersion regime. The wavelengths are the telecom wavelengths 1550 nm and 1310nm. The inset presents the cross-section profiles of the two beams after a propagation distance of $10 mm$.}

\FloatBarrier

\par{An important \textbf{remark}: from theoretical point of view, as the beams have different wavelengths, the exterior input angles ($\theta=\arcsin(\lambda/4s)$) needed to excite the quasi-nondivergent beams (i.e. at $k_x=\pi/2$) are different ($\theta_{1310} \neq \theta_{1550}$). But in our numerical Runge-Kutta simulations we used the same input angle for the both beams, i.e. the average angle $\theta = (\theta_{1310}+\theta_{1550})/2$. The reason for doing so is to be as close as possible to experimental conditions, where, in the many cases, the two beams overlap and have the same propagation direction, thus entering the same angle in the dispersive device.}

\par{As one can see, after 1cm propagation distance the two beams are well separated. From the inset, one can also see the $\lambda=1550 nm$ beam has the maximum amplitude a little bit smaller but the waist a little bit larger than the $\lambda=1310 nm$ beam. The cause is the greater influence of the discrete diffraction at $\lambda=1550 nm$, as the input beams have finite waists (instead of infinite ones as the theory requires) and the input angles are different from the ones needed to excite the non-divergent beams ($k_x=\pi/2$).}

\par{The coupling values calculated on figure \ref{fig2} can be used for estimates of the size of more conventional 2 guide couplers. We find 1310 nm et 1550 nm wavelengths can be separated with few millimeter long couplers. But in contrast with the waveguide array demonstrator discussed above, accurate design and fabrication of the interguide coupling is required for couplers. Sellmeier formulas yield a angular dispersion for niobate bulk material of the same order of magnitude as the waveguide arrays but the array arrangement provides flexibility with new regions of achromatic and anomalous dispersion.}

\section{Conclusion}

In this work we investigated the dispersive properties of a LN waveguide array in near-infrared wavelength range.

\par{The main conclusion we drawn is that decreasing the period, one passes from normal array dispersion where beam's slope increases with wavelength by achromatic point, to anomalous array dispersion where beam's slope decreases with wavelength. The solidity of the result is enhanced by the fact that both approaches, i.e. coarse and accurate evaluations yield the same behavior.}

\par{More important, dispersive properties can be tuned by modifying the period. This aspect, together with the fact that the values of all the parameters (waveguides separation, core-cladding index contrast, index profile, propagation distance, etc...) fall within ranges that are technological feasible according to the literature on LN waveguides fabrication and characterization, open new possibilities for optical data processing.}

\section*{Acknowledgments}
Authors are grateful to "INQCA - Integrated Quantum Circuits based on non-linear waveguide Arrays" Romania-France joint project PN-II-ID-JRP-RO-FR-2014-0013".


\begin{thebibliography}{99}

\bibitem{1}J. Moison, N. Belabas, C. Minot and J. A. Levenson, ``Discrete photonics in waveguide arrays," \ol \textbf{34}(16), 2462--2464 (2009).
\bibitem{2}N. Belabas, C. Minot, G. Bouwmans, J. A. Levenson and J. M. Moison, ``Discrete photonics resonator in coupled waveguide arrays," \opex {\bf 22}(10), 12379--12391 (2014).
\bibitem{3}R. Morandotti, H. Eisenberg, Y. Silberberg, M. Sorel and J. Aitchison, ``Self-focusing and defocusing in waveguide arrays," \prl {\bf 86}(15), 3296--3299 (2001).
\bibitem{4}D. N. Christodoulides and E. D. Eugenieva, ``Blocking and routing discrete solitons in two-dimensional networks of nonlinear waveguide arrays," \prl {\bf 87}(23), 233901 (2001).
\bibitem{5}J. Meier, G. I. Stegeman, D. N. Christodoulides, R. Morandotti, G. Salamo, H. Yang, M. Sorel, Y. Silberberg and J. S. Aitchison, ``Incoherent blocker soliton interactions in Kerr waveguide arrays," \ol \textbf{30}(23), 3174--3176 (2005).
\bibitem{6}D. Castaldini, P. Bassi, P. Aschieri, S. Tascu, M. De Micheli, P. Baldi, ``High performance mode adapters based on segmented SPE:$LiNbO_3$ waveguides," \opex {\bf 17}(20), 17868--17873 (2009).
\bibitem{7}D. Castaldini, P. Bassi, S. Tascu, P. Aschieri, M. De Micheli, P. Baldi, ``Soft proton exchange tapers for low insertion loss $LiNbO_3$ devices," Journal of Lightwave Technology, {\bf 25}(6), 1588--1593, (2007).
\bibitem{8}R. Iwanow, R. Schiek, G. Stegeman, T. Pertsch, F. Lederer, Y. Min and W. Sohler, ``Arrays of weakly coupled, periodically poled lithium niobate waveguides: beam propagation and discrete spatial quadratic solitons," Optoelectronic review {\bf 13}(2), 113 (2005).
\bibitem{9}J.H Zhao, X.H. Liu, Q. Huang, Peng Liu, Lei Wanga and Xue-Lin Wanga, ``The array waveguides formed in $LiNbO_3$ crystal by oxygen-ion implantation," Nuclear Instruments and Methods in Physics Research B {\bf 268}(19), 2923--2925 (2010).
\bibitem{10}H. Chen, T. Lv, A. Zheng and Y. Han, ``Discrete diffraction based on electro-optic effect in periodically poled lithium niobate," Optics Communications  {\bf 294}, 202--207 (2013).
\bibitem{11}A. Kaplan and S. Ruschin, ``Optical switching and power control in $LiNbO_3$ coupled waveguide arrays," IEEE Journal of Quantum Electronics {\bf 37}(12), 1562--1573 (2001).
\bibitem{12}R. Kruse, F Katzschmann, A Christ, A Schreiber, S Wilhelm, K Laiho, A G´abris, C S Hamilton, I Jex and C Silberhorn, ``Spatio-spectral characteristics of parametric down-conversion in waveguide arrays," New Journal of Physics {\bf 15}, 083046 (2013).
\bibitem{13}J.M. Moison, N. Belabas, J.A. Levenson and C. Minot, ``Light-propagation management in coupled waveguide arrays: quantitative experimental and theoretical assessment from band structures to functional patterns," \pra {\bf 86}, 033811 (2012).
\bibitem{14}C. Minot, N. Belabas, J. A. Levenson and J. M. Moison, ``Analytical first-order extension of coupled-mode theory for waveguide arrays," \opex {\bf 18}(7), 7157--7172 (2010).
\bibitem{15}L. Chanvillard, P. Aschiéri, P. Baldi, D.B. Ostrowsky, M.P. de Micheli, L. Huang, D.J. Bamford, ``Soft proton exchange on periodically poled $LiNbO_3$: A simple waveguide fabrication process for highly efficient nonlinear interactions," \apl {\bf 76}(9), 1089--1091 (2000).
\bibitem{16}David E. Zelmon, David L. Small and D. Jundt, ``Infrared corrected Sellmeier coefficients for congruently grown lithium niobate and 5 mol.\% magnesium oxide–doped lithium niobate," \josab {\bf 14}(12), 3319--3322 (1997).
\bibitem{17}G. Renversez, B. Kuhlmey and R. McPhedran, ``Dispersion management with microstructured optical fibers: ultraflattened chromatic dispersion with low losses," \ol \textbf{28}(12), 989-991 (2003).
\bibitem{18}A. Hartung, A.M. Heidt, and H. Bartelt, ``Design of all-normal dispersion microstructured optical fibers for pulse-preserving supercontinuum generation," \opex {\bf 19}(8), 7742--7749 (2011).
\bibitem{19}A.M. Apetrei, M.C.P. Huy, N. Belabas, J.A. Levenson, J.M. Moison, J.M. Dudley, G. Mélin, A. Fleureau, L. Galkovsky, S. Lempereur, ``A dense array of small coupled waveguides in fiber technology: trefoil channels of microstructured optical fibers," \opex {\bf 16}(25), 20648-20655 (2008).
\bibitem{20}A.M Apetrei, J.M. Moison, J.A. Levenson, M. Foroni, F. Poli, A. Cucinotta, S. Selleri, M. Legré, M. Wegmüller, N. Gisin, K.V. Dukel'Skii, A.V. Khokhlov, V.S. Shevandin, Y.N. Kondrat'Ev, C. Sibilia, E.E. Serebryannikov, A.M. Zheltikov, ``Electromagnetic field confined and tailored with a few air holes in a photonic-crystal fiber," Applied Physics: B-Lasers and Optics {\bf 81}(2-3), 409-414 (2005).


\end{thebibliography}
\end{document}